\documentclass[12pt]{article}

\begin{document}
\topmargin 0pt
\oddsidemargin 0mm

\renewcommand{\thefootnote}{\fnsymbol{footnote}}
\begin{titlepage}
\begin{flushright}
IC/2001/59 \\
IP/BBSR/2001-15\\
hep-th/0106232
\end{flushright}

\vspace{5mm}
\begin{center}
{\Large \bf Noncommutativity in linear dilaton background}
\vspace{10mm}

{\large
Sandip Bhattacharyya$^{**}$, Alok Kumar$^{**}$ and 
Swapna Mahapatra$^{*, \dag}$\\
\vspace{8mm}
$^{**}${\em Institute of Physics, Bhubaneswar 751 005, India}\\

\footnote{Present address, Regular Associate of the Abdus 
Salam ICTP}{\em Abdus Salam International Centre for Theoretical
Physics, \\
 Strada Costiera 11, 34014 Trieste, Italy \\
and \\
\footnote{Permanent address}Physics Department, Utkal University, 
Bhubaneswar 751 004, India}\\ 

\vspace{5mm}

email: sandip,kumar,swapna@iopb.res.in}

\end{center}
\vspace{5mm}
\centerline{{\bf{Abstract}}}
\vspace{5mm}

We consider quantization of open string theories in 
linear dilaton and constant antisymmetric tensor
backgrounds and discuss the noncommutativity of space-time 
coordinates arising in such theories, including their 
relationship with light-like noncommutativity as well as 
backgrounds with null isometries. 
It is argued that the results can also be understood using 
space-time equations of motion of the string modes. We then present 
$N=2$ supersymmetric generalization of 
these theories and the associated noncommutative
structure. 

\end{titlepage}

\newpage

The study of string propagation in background fields has been an active
area of research for a long time \cite{friedan}. One such consistent 
set of background includes a general target space metric, antisymmetric 
tensor field and a linear dilaton. String theories with linear dilaton 
background \cite{myers} are interesting from the point of view of 
$D$-branes \cite{li,rajaraman}, as well as conformal 
field theory (CFT) using their free field realizations. 
They also play a role in understanding the cosmological implications  
of string theory\cite{bachas} and their applications to 
quantum gravity, through the constructions of 
noncritical string theories \cite{sasha,marn,david,DK,nati}, 
where the role of 2-dimensional gravity  becomes important. 
Thus such theories are important to study, in order to have a better
understanding of gravity in general, and also in constructing interesting 
physical systems with nontrivial 
quantum gravity effects. Recently, D-branes in linear dilaton background
have been studied, where the corresponding Dirichlet boundary state 
has been constructed \cite{li,rajaraman}. 
The aim of this paper is to further study the linear 
dilaton background in the context of D-branes. In particular, our aim 
is to analyze the non-commutative structure \cite{SW} 
arising in such background when a constant antisymmetric tensor field
is turned on along these D-branes.

We now consider open strings in $d$ dimensional space-time in the 
presence of a constant Neveu-Schwarz antisymmetric tensor field
$B$ and a dilaton $\Phi$. The world sheet action is given by 
\cite{callan},

\begin{eqnarray}
S & = & \frac{1}{4\pi\alpha'} \int_{\Sigma} d^2 z 
\big [{\sqrt \gamma}{\gamma}^{a b} 
\partial_{a}X^{\mu}\partial_{b}
X^{\nu} g_{\mu\nu} + \epsilon^{a b}\partial_{a}
X^{\mu}\partial_{b}X^{\nu} B_{\mu\nu} - \frac{1}{2}
\alpha'{\sqrt \gamma}R^{(2)}\Phi(X) \big ] - \nonumber \\
& & \frac{1}{4\pi\alpha'}\oint_{\partial\Sigma}d\tau 
\big [\alpha' k\Phi(X) \big ].
\end{eqnarray}
Here we have taken the background gauge fields to be zero. 
$R^{(2)}$ is the two-dimensional curvature scalar and $k$ is 
the extrinsic 
curvature of the boundary. We take the world sheet  metric 
to be Minkowskian and $\epsilon^{01} = 1$. The $d$ dimensional 
index $\mu$ runs from $0, \ldots, d-1$. This world sheet action 
in conformal gauge reduces to,
\begin{equation}
S_0 = {1\over 4 \pi \alpha'} \int d^2 z \left[  
g_{\mu \nu} \partial_a X^{\mu} \partial^a X^{\nu} 
+  \epsilon^{a b} B_{\mu \nu} \partial_a X^{\mu} \partial_b X^{\nu} 
+ {\alpha'} \partial_a \rho \nabla_{\mu} \Phi \partial^a X^{\mu}
\right].
\label{action1}
\end{equation}
where, $\rho$ is the conformal mode of the metric. We assume that 
the dilaton $\Phi$ is of the form
$\Phi = Q_{\mu} X^{\mu}$, i.e., a linear function of the 
coordinates where $Q$ is the background charge.  One also takes  
$Q^2 \equiv Q_{\mu} Q^{\mu} = {(26 - d)\over 24\alpha'}$, in order to 
cancel the conformal anomaly in the quantized theory. 
We will further restrict ourselves to the gauge $\rho = 0$.
One can then directly apply the
procedure of \cite{chu-ho} to quantize the theory. 
Equations of motion, boundary conditions, as well as the canonical 
momenta remain unchanged with respect to the case when 
$Q_{\mu}=0$ and $d=26$. However, since the energy-momentum tensor 
acquires an extra term due to the coupling of the dilaton to the 
worldsheet curvature scalar, the Virasoro generators are also 
modified accordingly with terms depending on the background charge 
$Q$. 
 
The variation of the action gives the boundary conditions at $\sigma = 0, 
\pi$ as 
\begin{equation}
\partial_{\sigma} X^{\mu} + F_{\nu}^{\mu} \partial_{\tau} X^{\nu}|_
{\sigma = 0, \pi} = 0,
\label{boundary}
\end{equation}
and from the equations of motion, the general solutions for the string 
coordinates $X^{\mu}$ are obtained as, 
\begin{equation}
X^{\mu} = x_0^{\mu} + (p_0^{\mu} \tau - p_0^{\nu} F_{\nu}^{\mu} \sigma) 
+ \sum_{n\neq 0}{e^{- i n \tau}\over n}
(i a^{\mu}_n \cos n\sigma - a^{\nu}_n F_{\nu}^{\mu} \sin n\sigma).
\label{mode}
\end{equation}
In flat Minkowski space ($g_{\mu \nu} = \eta_{\mu \nu}$), one has 
$F_{\mu}^{\nu} \equiv B_{\mu}^{\nu}$. 
Also, the mode expansion for the canonical momenta 
can be written in a form: 
\begin{equation}
2\pi\alpha' P^{\mu}(\tau, \sigma) = G^{\mu}_{\nu}
(a_0^{\nu} + \sum_{n \neq 0} a_n^{\nu} e^{-i n \tau}
\cos n\sigma ),
\end{equation}
with $G$ being the open string metric given by, 
\begin{equation}
G = g - B g^{-1} B.
\end{equation}

The noncommutative properties and the canonical commutation 
relations in this theory are identical to the ones in 
\cite{chu-ho} and are described by the following relations :
\begin{eqnarray}
\left [x_0^{\mu}, x_0^{\nu} \right ] & = & i \pi \alpha' 
\left [ (G^{-1})^{\mu \rho} F_{\rho}^{\nu} - (G^{-1})^{\nu \rho} 
F_{\rho}^{\mu} \right ], \nonumber \\
\left [ x^{\mu}_0, p^{\nu}_0 \right ] & = & 2 i \alpha' 
(G^{-1})^{\mu\nu}, \nonumber \\
\left [ a^{\mu}_m, a^{\nu}_n \right ] & = & 2 \alpha' m (G^{-1})^{\mu \nu} 
{\delta}_{m + n, 0} \nonumber \\
\left [ a^{\mu}_m, x^{\nu}_0 \right ]  & = & 0, \nonumber \\ 
\left [a^{\mu}_m, p^{\nu}_0 \right ]  & = & 0. 
\label{commute}
\end{eqnarray}

To write down the Virasoro generators, one starts with the expression 
for the components of the worldsheet energy-momentum tensor. 
It is useful to express quantities in complex coordinates $z$ and 
$\bar z$, where the world sheet of the open string theory corresponds 
to the upper half plane, Im $z > 0$. The energy momentum tensor
can then be written as, 
\begin{eqnarray}
T_{zz} & = & {1\over 4 \alpha'} g_{\mu\nu} {\partial X}^{\mu}
{\partial X}^{\nu} + Q_{\mu}{\partial}^2 X^{\mu}, \nonumber \\
T_{\bar z \bar z} & = & {1\over 4\alpha'} g_{\mu\nu} \bar\partial
X^{\mu} \bar\partial X^{\nu} + Q_{\mu}{\bar\partial}^2 X^{\mu}.
\label{t++--}
\end{eqnarray}
Using the mode expansions for the coordinates as in equation 
(\ref{mode}), one can expand $T_{z z}$, $T_{\bar{z} \bar{z}}$ as:
\begin{eqnarray}
 T_{z z} & = & \sum_m 
\left({1\over 4\alpha'} 
        \sum_n G_{\mu \nu} a^{\mu}_{m-n} a^{\nu}_n + 
        i (m +1) Q^{\mu} (g + B )_{\mu \nu} a^{\nu}_{m} \right)
         {z}^{-m - 2}, \nonumber \\
T_{\bar{z} \bar{z}} & = & \sum_m \left( {1\over 4\alpha'} 
         \sum_n G_{\mu \nu} a^{\mu}_{m-n} a^{\nu}_n + 
        i (m + 1) Q^{\mu} (g - B )_{\mu \nu} a^{\nu}_{m}
         \right) {\bar{z}}^{-m - 2}.
\label{t-mode}
\end{eqnarray}
One can also express them in an alternative form of the mode 
expansion, where the $Q$-dependent terms appear with coefficient 
$m$ rather than $(m+1)$. They can, however, be obtained from 
the above expressions by a shift proportional to $Q^{\mu}$ 
on $\alpha_0^{\mu}$.
 
Moreover, in open string theories, the conformal invariance condition 
on the boundary implies, a matching 
condition, $T_{z z} = T_{\bar{z} \bar{z}} |_{z = \bar{z}}$ ~(Im $z = 0$), 
which is
derived from the requirement that there is no net flow of 
energy and momentum from the boundary.
In our case, this constraint 
is satisfied by imposing the condition 
\begin{equation}
Q^{\mu} B_{\mu \nu} = 0.
\label{condition}
\end{equation} 
Also, using relations (\ref{commute}), 
it can be verified that $T_{z z}$ as well as $T_{\bar{z} \bar{z}}$ 
satisfy conformal algebra with central charge 
$C= d + 24 \alpha' Q^2$. 

There are several solutions for the above condition between the 
background charges $Q_{\mu}$'s and the antisymmetric tensor 
components $B_{\mu \nu}$'s.  First, if the directions along which 
$Q_{\mu}$'s are turned on, are orthogonal to the ones which have 
constant antisymmetric tensor couplings, conditions (\ref{condition}) are
trivially satisfied.  More interesting solutions arise 
when the electric and magnetic components of the $B$ field are related as 
in the case of light-like noncommutativity \cite{aharony}.
The above condition on the background charge then implies that 
$Q_{\mu}$ is also a light-like vector. 
Theories with light-like noncommutativity are known to be 
unitary \cite{gomis} and have a nice decoupling limit. 
In particular, the light-like noncommutativity condition reads 
$B_{0 i} = B_{1 i} \neq 0$. Noncommutativity with parallel electric 
and magnetic field components has also been studied in ref. 
\cite{jabbari}. Let us just consider the case where 
$B_{12}$ as well as $B_{02}$ components are nonzero. 
Then the light-like noncommutativity condition, 
eqn. (\ref{condition}), implies that the two corresponding
background charges satisfy the condition $Q_0 = Q_1$.
The dilaton is then proportional to both the time as well
as space coordinates, ${\it i. e.}$ $\Phi = Q_0 (X^0 + X^1)$.
Similar type of dilaton backgrounds have been considered before
in \cite{bershadsky}. They also appear in asymptotic limits 
of space-time backgrounds in various gauged WZW models.
In our case, they belong to the  class of 
space-time backgrounds with null isometries\cite{chiral}, and 
are known to be classical solutions of string theories to all 
orders in $\alpha'$. 
 
It remains to be seen whether the condition (\ref{condition})
is also a necessary one to define a consistent quantum theory. 
In other words, it may be interesting to understand
whether the matching condition at the boundary given above
can be treated as a constraint in the quantization process. 
In that case, equation (\ref{condition}) will be replaced by an  
appropriate physical state conditions.

We also note that the mode expansions for $T_{z z}$ and 
$T_{\bar{z} \bar{z}}$,
given above, are related by $B \rightarrow - B$
transformation. Such asymmetry in the form of fields and 
operators occurs in open string theory, even in the constant dilaton case. 
For example, in the case of NSR superstring in constant $B$ 
background, the mode expansions of left and right-handed 
Neveu-Schwarz and Ramond fermion fields are obtained as \cite{itoyama},
\begin{eqnarray}
\psi^i (z) & = & \sum_{r} \left [g^{-1} (g - 2\pi \alpha' B) \right ]^i_j
b_r^j z^{- r - {1\over 2}}, \nonumber \\
\bar{{\psi}^i}(\bar z) & = & \sum_{r} \left [g^{-1} 
(g+ 2\pi \alpha' B) \right ]^i_j b_r^j {\bar z}^{- r- {1\over 2}},   
\label{psi-psi}
\end{eqnarray}
where $r$ is an integer (half-integer) in R-sector 
(NS-sector). Operators 
$\partial X^{\mu}$, $\bar{\partial}X^{\mu}$ also have similar 
mode expansions\cite{itoyama} and correspond to 
the worldsheet conserved currents for the translations
in the left and the right-moving sector.
The worldsheet fields $\psi$, $\bar{\psi}$ etc. 
appear in  vertex operators for vector fields
in superstring theory \cite{itoyama} as a linear 
combination $(\psi^{\mu} + \bar{\psi}^{\mu}) e^{i k.X}$  
of the left and right-moving components.  
For the case of Virasoro generators, however, there is an extra 
condition on the form of $B$'s and $Q_{\mu}$'s originating from 
the matching condition on the boundary. 
One can then use the doubling
trick where $T_{zz}$ becomes holomorphic in the whole complex plane 
and there is then one set of Virasoro generators, given by the 
coefficients of the mode expansions in equation (\ref{t-mode}).

We now discuss the existence of consistent open string theory in  
constant $B$ and linear dilaton backround by using certain space-time 
equations of motion. In particular, the equation of motion for the 
tachyon in the open string sector can be written 
in these backgrounds as \cite{nappi,callan,friedan},
\begin{equation}
{(g - B g^{-1} B)}^{-1}_{\mu \nu} \nabla^{\mu}\nabla^{\nu}
\Theta - \nabla^{\mu}\Phi \nabla_{\mu} \Theta + 
{1\over \alpha'} \Theta = 0. 
\label{o-tach}
\end{equation}
This is the tachyon $\beta$-function equation obtained by 
considering string theories in nontrivial 
backgrounds\cite{callan,friedan}.
However, one notices that for constant $g$, $B$ and a linear dilaton
$\Phi$, this equation can be reinterpreted in terms of  a 
modification to the energy spectrum.
For example, assuming a form like $\Theta = e^{i k\cdot x}$ for the 
open string tachyon wave function,
one obtains a condition from equation (\ref{o-tach}) such as,
\begin{equation}
G^{\mu \nu} k_{\mu} k_{\nu} + i k_{\mu} Q^{\mu}
- {1\over \alpha'} =0, 
\label{weight-o}
\end{equation}
with $G = g - B g^{-1} B$ being the open  string metric. 
The left hand side in equation (\ref{weight-o})
can be identified with 
the conformal weight that appears using the Virasoro generators
we obtained above and confirms the validity of our construction. 

To continue with the analogy, we notice that the equation of motion for 
the gauge field `fluctuations' in the background of constant $g$ and
$B$, and linear dilaton can be written (by dropping interaction terms)
in the form\cite{callan,friedan}, 

\begin{equation}
{(g - B g^{-1} B)}^{-1}_{\lambda \nu}\nabla^{\nu} f_{\mu}^{\lambda}
+ \nabla^{\nu}\Phi f_{\nu \mu} =  0,
\label{o-gauge}
\end{equation}
provided the condition (\ref{condition}) once again holds. 
Now, assuming the wave function for the gauge 
field fluctuation to be of the form:
$a_{\mu} = \epsilon_{\mu} e^{i k\cdot X}$,
we obtain the mass-shell condition as,

\begin{equation}
 \tilde{h} = G^{\mu \nu} k_{\mu} k_{\nu} + i k^{\mu} Q_{\mu} = 0
\label{weight-photon}
\end{equation}
and the physical state condition reads as,

\begin{equation}
(k_{\mu} G^{\mu \nu} + Q^{\nu} ) \epsilon_{\nu} = 0.
\label{physical}
\end{equation}

Using the expressions of the conformal generators mentioned above, 
we then reproduce the expressions for the 
conformal weight ($\tilde{h}$) in (\ref{weight-photon}), 
as well as the physical 
state condition (\ref{physical}). We find it interesting to 
observe that the condition (\ref{condition}) appears from 
the point of 
view of CFT, as well as in the study of the space-time equations of
motion. We also mention that, although the above set of 
equations, (\ref{o-tach})-(\ref{physical}), 
for string modes is derived from the string effective action 
in commutative quantization scheme, one can argue that by 
restricting to the
linearized fluctuations and ignoring interaction terms, one gets 
identical conditions in noncommutative framework as well. 

We now discuss $N=2$ (worldsheet) supersymmetric\cite{petersen,zheng} 
extension of our
results in the critical string theory case.   
This analysis is of importance for  studying the supersymmetric 
version of D-branes with linear dilaton background as well as
in their constructions using $N=2$ minimal models \cite{reck}.  
To present such an extension of the above analysis,
we now give the representation of $N=2$ superconformal field 
theories using free bosonic and fermionic fields having  
mode expansions of the type (\ref{mode}), (\ref{psi-psi}),
and superconformal generators are the generalizations of the
ones appearing in  (\ref{t++--}, \ref{t-mode}).
We write down explicit expressions for the superconformal generators 
and show that they satisfy the $N=2$ operator algebra. 
To be specific, in this part of the paper, we first restrict
ourselves to the case when $g$, $B$ etc.are $2\times 2$ matrices. 
We then start with the 2-point function 
for bosons $\phi^i$ and 
fermions $\psi^i$, $(i=1, 2)$ that can be obtained by 
turning on constant $B$. By restricting ourselves to the 
boundary, these 2-point functions are:
\begin{eqnarray}
\langle \phi^i(\tau_1)\phi^j(\tau_2) \rangle  & = &
- G^{i j} ln (\tau_1 - \tau_2)^2
                 + \theta^{ i j} \epsilon(\tau_1 - \tau_2),\nonumber \\
\langle \psi^i(\tau_1) \psi^j(\tau_2) \rangle & = & - {G^{i j}\over 
{(\tau_1 - \tau_2)}}. 
\end{eqnarray}
with $G = g  - B g^{-1} B$, $g$ now being a two dimensional identity 
matrix
and $B_{i j} = b \epsilon_{i j}$. $\theta_{i j}$ are the 
noncommutativity
parameters: $\theta_{i j} = \theta \epsilon_{i j}$, with 
$\theta = - {b\over {1+b^2}}$. 
Now following \cite{zheng}, we work in a complex notation, where,
\begin{eqnarray}
A = \phi^1 + i \phi^2, ~~~~~ \bar{A} = \phi^1 - i \phi^2, \cr
\psi = \psi^1 - i \psi^2, ~~~~~ \bar{\psi} = \psi^1 + i \psi^2.
\label{a-bar-a}
\end{eqnarray}
The boundary two point functions are then given by, 
 
\begin{eqnarray}
\langle A(\tau_1) \bar{A}(\tau_2) \rangle  & = 
& -{2\over 1 + b^2} ln(\tau_1 - \tau_2)^2
  + \theta \epsilon(\tau_1 - \tau_2), \nonumber \\
\langle \bar{A}(\tau_1) A(\tau_2) \rangle & = 
& -{2\over 1 + b^2} ln(\tau_1 - \tau_2)^2
                           - \theta \epsilon(\tau_1 - \tau_2), 
\label{AbarA}
\end{eqnarray}
and 
\begin{equation}  
\langle \psi(\tau_1) \bar{\psi}(\tau_2) \rangle  ~ = ~ -{1\over (1+ b^2)(\tau_1
  -\tau_2)} ~ = ~ \langle \bar{\psi}(\tau_1) \psi(\tau_2) \rangle .
\label{two point function}
\end{equation}
We now write down the expression for the superconformal generators,
by extending the expressions for the conformal generators written 
earlier in equation (\ref{t++--}), (\ref{t-mode}) 
to the full $N=2$ superconformal symmetry. 
We give these expressions in terms of 
conformal fields, rather than their oscillator modes. They can,
however, be mapped into each other. The full set of generators
in $N=2$ theories consist of the $U(1)$ generators $J$, two 
superconformal generators $G, \bar{G}$ and the conformal generator
$T$. Their conformal weights are $1, {3\over 2}$ and 
$2$ respectively. In our case, they have a form:
\begin{eqnarray}
J & = & {1\over 2}[(1+b^2){\bar{\psi}\psi} + (1+ i b)\bar{\beta}
\partial_{\tau} A - 
(1- i b)\beta\partial_{\tau} \bar{A}], \nonumber \\
G & = & {i\over \sqrt{2}} (1 + b^2) \psi\partial_{\tau} A - 
 2\sqrt{2} i \beta (1- i b)
 \partial_{\tau} \psi, \nonumber \\
\bar{G} & = & - {i\over \sqrt{2}} (1+b^2) \bar{\psi}\partial_{\tau} 
\bar{A} + 2\sqrt{2} i \bar{\beta} (1+ i b)
\partial_{\tau} \bar{\psi},\nonumber \\
T & = & -{1\over 4} (1+ b^2)\partial_{\tau} A 
          \partial_{\tau} {\bar{A}} 
         - {1\over 2} (1 + b^2) \partial_{\tau}\bar{\psi}\psi 
         + {1\over 2} (1 + b^2)\bar{\psi}\partial_{\tau}\psi + \nonumber \\
& & {\bar{\beta}\over 2}(1+ i b)\partial_{\tau}^2 A
         +{{\beta}\over 2}(1 - i b)\partial_{\tau}^2 \bar{A}.
\label{n2generators}
\end{eqnarray}
Here $\beta$ and $\bar{\beta}$ are complex parameters. 
By writing down 
$A$, $\bar{A}$ etc. using oscillator mode expansions, such as 
the ones in (\ref{mode}), and restricting to the boundary 
at $z = \bar z$, 
one obtains the mode expansions of the 
above generators which are the generalizations of the ones appearing
in equation (\ref{t++--}, \ref{t-mode}) to the supersymmetric case.  
To verify that operators in equation 
(\ref{n2generators}) satisfy $N=2$  
algebra, one uses the two point functions in 
equations (\ref{AbarA}), (\ref{two point function}). 
The parameter $\theta$, appearing in these expressions  
does not play a 
role in verifying the $N=2$ superconformal algebra which involves a time 
ordering of the operators on the boundary. Moreover, since 
Wick contraction in this context always involves at least one
$\tau$ derivative, this term drops out from the computation.

We also mention that the generators in equation 
(\ref{n2generators}) have been obtained from the reduction of the 
holomorphic part of the superconformal generators to the 
boundary. As mentioned before, 
if one starts with the antiholomorphic generators, 
one gets expressions which are related to the ones in 
(\ref{n2generators}) by a transformation: $B \rightarrow - B$. 

In fact this construction of $N=2$ generators can be extended to 
several copies of bosonic fields $A^I, \bar{A}^{\bar{I}}$ and
fermionic fields $\psi^{\bar{I}}, \bar{\psi}^{I}$
\footnote{the indices on $\psi$ and $\bar{\psi}$ correspond
to their definitions in equation (\ref{a-bar-a})}. The 
non-vanishing components of the (closed string) metric and 
antisymmetric tensors are: $g_{I \bar{J}}$ and $B_{I \bar{J}}$. 
$N=2$ generators can be written for these cases by generalizing the 
expressions in equation (\ref{n2generators}) to include indices 
$I, \bar{I}$ over the fermions and bosons. At the same time 
the expression $(1+ b^2)$ and  $(1 \pm i b)$ 
appearing in equation (\ref{n2generators})
are replaced by : $G_{{I} \bar{J}}$ and  
$(g_{I \bar{J}} \pm i B_{I \bar{J}})$ respectively, with $G_{I \bar{J}}$,
now being the `open' string metric. The matching of the $N=2$ 
generators on the boundary once again gives a condition
involving $B$, and  background charges $\beta, \bar{\beta}$, 
representing the  linear dilaton in the present case:
$\beta^I B_{I \bar{J}} = 0$,
$\bar{\beta}^{\bar{I}} B_{\bar{I} J} = 0$.

We have therefore given the constructions of 
noncommutative open string theories in linear 
dilaton and constant antisymmetric tensor backgrounds.
We also note that the classical action (\ref{action1}) 
discussed above, appears in general in the context of noncritical string 
theories\cite{sasha,marn,das,das2}, where 
2-d gravity couples to the matter. In view of our analysis, it 
will be interesting to investigate non-commutativity in these cases
as well. We end with a comment that the relationship 
with light-like noncommutativity, as found above,   
deserves further study. In particular, one would also like to 
understand the decoupled field theory limit \cite{aharony, jabbari,
chen, cai} in the above case and 
the corresponding dual supergravity descriptions, as has been 
discussed in other theories with light-like noncommutativity 
\cite{alishahiha}.  
It may also be interesting to understand our results in the context of 
little string theories \cite{berkooz} where the dual theory is a 
theory without gravity in lower dimension.

{\bf Acknowledgement: } We thank M. M. Sheikh-Jabbari, A. Misra,
S. Mukherji and K. Ray for many useful discussions. 
S. M. would like to thank the Abdus Salam I.C.T.P. for an 
Associateship under which this work was done.


\end{document}